# Magnetic Dipoles and Quantum Coherence in Muscle Contraction

Kuniyuki Hatori, Hajime Honda and Koichiro Matsuno[*]

Department of BioEngineering, Nagaoka University of Technology,
Nagaoka 940-2188, Japan

**An actin filament contacting myosin molecules as a functional unit of muscle contraction induces magnetic dipoles along the filament when ATP molecules to be hydrolyzed are available there. The induced magnetic dipoles are coherent over the entire filament, though they are fluctuating altogether as constantly being subject to the ambient thermal agitations.**

[*]To whom correspondence should be addressed. E-mail: kmatsuno@vos.nagaokaut.ac.jp

There have been both experimental and theoretical suggestions on the likelihood of mesoscopic coherence of quantum mechanical origin in various biological phenomena [1, 2]. In fact, quantum dipole couplings of entangled nuclear spins in two molecules separated by distances even over one millimeter are detectable as a measurable NMR signal [3]. Quantum coherence and superposition [4, 5] could also be expected from the coherent flow of electrons [6] and of protons [7-10]. One demonstrative process actualizing proton currents is an ATP hydrolysis in an actomyosin complex underlying muscle contraction. We examine the nature and the phase characteristic of magnetic dipoles to be induced along an actin filament contacting myosin molecules in the presence of ATP molecules to be hydrolyzed.

    Our primary objective was to measure the fluctuating displacements of an actin filament sliding on myosin molecules in the presence of ATP molecules under the influence of the magnetic flux applied externally. The secondary objective was to examine the nature of the magnetic dipoles induced along the filament by the flow of protons released from ATP molecules during their hydrolysis. For this purpose, we first prepared a speckled actin filament having several fluorescent markers along the filament [11]. Actin and myosin molecules were prepared from rabbit skeletal muscle. Speckled actin filaments were made from a mixture of actin filaments, both unlabeled and labeled by rhodamine-phalloidine that is a fluorescent material. Unlabeled and labeled actin filaments [25 mM KCl, 25 mM imidazole-HCl (pH=7.4), 4 mM DTT] were made and treated with equal molar phalloidine, and their concentrations were maintained at 33 µg/ml. The mixture of the suspensions of labeled and unlabeled actin filaments with a volume ratio of 1:4 was subjected to supersonic dissecting (Misonix, Microson Model XL2000) for 4 min. The mixture of dissected actin filaments, both unlabeled and labeled mixed together, was placed at temperature 4°C for 24h for reformation of the filaments that could be speckled as displayed in Figure 1.

    We then prepared a standardized *in vitro* motility assay [12]. The slide glass to fix

myosin molecules of 0.05mg/ml HMM (heavy meromyosin) was hydrophobically treated with nitrocellulose (Shin-etsu Chem.). The solution condition for observing the sliding movement of a speckled actin filament was 25 mM KCl, 25 mM imidazole-HCl (pH=7.4), 4 mM KCl, 1 mM ATP, 1 mM DTT at temperature 25$^o$C. The microscope (Olympus, IX70) attached with the object lens (Uplan Apo 100x, oil) was used with the aid of the accompanying fluorescent equipment and rhodamine filter. The observed images of actin filaments were taken and stored in a video casette recorder equipped with ICCD-camera (Video Scope, ICCD-350F). The images were processed with use of image-processing software NIH Image 1.6 (Wayne Rasband, NIH). Each image was taken at every 1/30 s through a video grabber board (Scion, LG-3 PCI) into a computer. The spacing of neighboring pixels was 60 nm, and the spatial resolution of identifying the center of each speckled segment along the filament was 20 nm.

We measured the trajectory of each center of the fluorescent marker, referred to as a reference point, attached on a speckled actin filament sliding on myosin molecules (HMM) with use of two independent methods. One was about the actual trajectory of each reference point on a planar plane at time $t$ updated every 1/30 s. The other was about the trajectory of each smoothed reference point at time $t$ updated every 1/30 s, that was defined as a spatially averaged point over the same actual reference points observed in the time interval [$t$-10/30 s, $t$+10/30 s] comprising 21 successive video frames. The trajectory of a smoothed reference point was in fact taken to be that of an actual reference point subjected to going through a low-pass filter, in which the high-frequency fluctuating components of the displacements, displaying staggered movements [13], had already been suppressed.

The displacement of an actual reference point relative to its corresponding smoothed reference point had two components. One was the parallel displacement tangential to the curvature of the trajectory of the smoothed reference point, and the other was the perpendicular displacement perpendicular to the curvature. Figure 2 displays the cross-correlation function of the fluctuating parallel displacements of two neighboring actual reference points measured relative to the corresponding smoothed reference points. The time interval required for estimating the time-averaged cross-correlation functions was chosen to be 3.3 s. Intensity of the cross-correlation at no time delay was found to increase as the magnetic flux density applied externally increased. The auto-correlation function of the parallel displacement of each actual reference point measured relative to the corresponding smoothed reference point, though not shown here, also demonstrated the similar enhancement with the increase of the applied magnetic flux. This enhancement manifests the occurrence of the fluctuating magnetic dipole moments induced along the actin filament, since magnetic dipole moment is energetically conjugate to magnetic flux. Intensity of the cross-correlation in the presence of magnetic flux was also found to remain almost unchanged even though the distance between the two points over which the cross-correlation was evaluated was increased up to the top-to-end of the entire filament. This relative invariance of the cross-correlation over varying distances reveals that the fluctuating strength of the induced magnetic dipole moment was almost in phase over the entire filament.

Increase in the fluctuating intensity of the parallel displacements of an actin filament in the presence of magnetic flux points to the internal tensile force generated there. When a magnetic dipole carrying its moment density $M$ [Ampere/meter] is put in the magnetic flux with its density $B$ [Tesla], the energy density $E$ [Joule/meter$^3$] of the dipole gives $E=-MB+B^2/2\mu_0$. Here, $\mu_0$ (=$4\pi\times10^{-7}$ [Tesla meter/Ampere]) is the magnetic permeability of the vacuum. The magnetic energy density $E$ becomes negative for $0<B<2\mu_0M$, implying that the magnetic dipole induces the internal tensile force with its strength ($MB-B^2/2\mu_0$) per unit area [Newton/meter$^2$]. The maximum tensile force $\mu_0M^2/2$ is expected at the magnetic flux density $B=\mu_0M$. Since the parallel displacements of an actin filament come to respond to the internal tensile force generated there, the maximal displacement can be expected at the magnetic flux density $B=\mu_0M$. Figure 3 demonstrates the relationship between the fluctuating intensity of the parallel displacements and the strength of the applied magnetic flux density. The maximal displacement was observed to occur at the flux density $B=65m$T [milli Tesla]. The magnetic dipole density was thus found to be $M=5.2\times10^4$A/m. Accordingly, the magnetic dipole moment per actin monomer was estimated to roughly be $1.7\times10^{-21}$Am$^2$ ($\cong 180\mu_B$, where $\mu_B$ is Bohr magneton). For this estimation, we assumed a coherent unit of magnetization to be an actin monomer of its diameter $4\times10^{-9}$m within which electrons forming covalent bonds are confined.

The magnetic dipoles induced along an actin filament exhibit a magnetic dipole-dipole interaction. This is seen in the relationship between the fluctuating intensity of the parallel displacements of an actin filament and the direction of the magnetic flux applied externally, as demonstrated in Figure 4. The fluctuating intensity increased as the direction along which an actin filament slid on myosin molecules maintained a certain angle against the direction of the magnetic flux applied externally.

Magnetic dipoles induced over an actin filament sliding on myosin molecules in the presence of ATP to be hydrolyzed were found to align with each other coherently over the entire filament though the strengths of their moments were fluctuating. Even in the presence of thermal agitations inducing rapid decoherence [14], the magnetic dipoles in ATP-activated actomyosin complexes can maintain their coherence over the entire actin filament [15-17]. The energy of the magnetic dipole-dipole interaction per actin monomer was about $1.1\times10^{-22}$ Joule which is far less than the thermal energy per degree of freedom available at room temperature. Nonetheless, the coherent stability in the alignment of the magnetic dipoles can be expected to be due to an anisotropic nature of the underlying dipole-dipole interaction. ATP hydrolysis at an actomyosin complex is energetically anisotropic in the sense that although the forward process of ATP hydrolysis is certainly possible in the presence of thermal agitations, the reverse process of making an ATP from an ADP and an inorganic phosphate is not available in the same thermal environment. Of course, detailed physical mechanisms implementing such a mesoscopic coherence of induced magnetic dipoles even in the presence of thermal agitations remain to be seen.

Thanks are due to Kazuyoshi Takahashi for experimental assistance and to Kazuhiko Shimada for support. One of the authors (K. M.) wishes to thank Michael Conrad, George L. Farre, Brian D. Josephson, Ray C. Paton, Karl H. Pribram and Walter J. Schempp for discussion at various phases of the work reported in this article.

**Figures Captions**

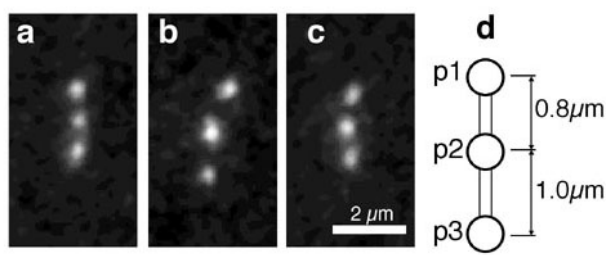

Figure 1: (a-c) Fluorescent image of a speckled actin filament. Three independent samples are displayed. (d) A schematic representation of a speckled actin filament. Circles correspond to fluorescent regions. These circles are denoted as *p1* through *p3*. The distances between the neighboring circles are measured as referring to the center of each circle.

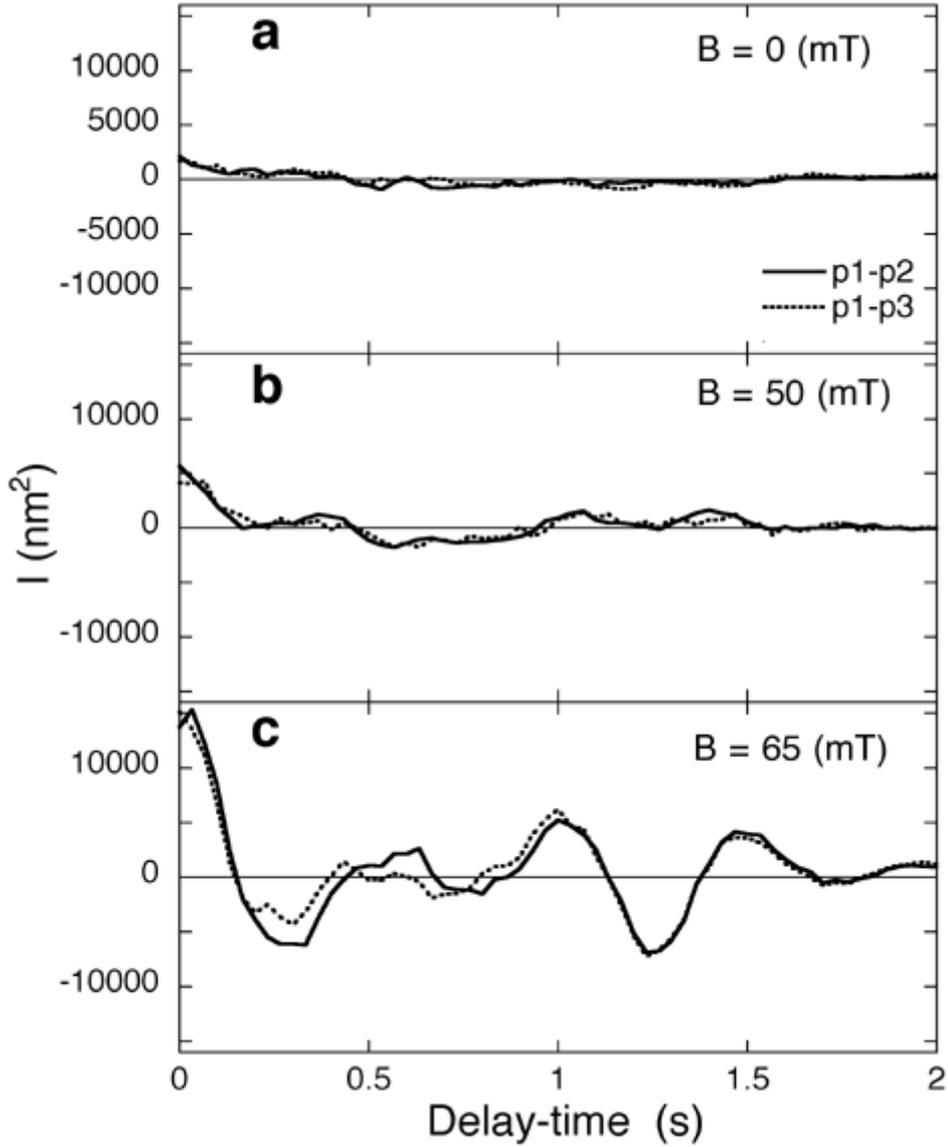

Figure 2: Intensity I (*nm*$^2$) of the cross-correlation function of the fluctuating parallel displacements between two actual reference points attached on a speckled actin filament parameterized in terms of the delay time for estimating the cross-correlation, and its dependence on the magnetic flux density B (*m*T) applied externally. Magnetic flux was applied in parallel to the planar plane on which a speckled actin filament slid. Two samples of the cross-correlation between *p1-p2* and between *p1-p3* are displayed. The actual reference points *p1, p2,* and *p3* follow the notation already displayed in Figure 1. The averaged distance between *p1* and *p2* was 0.8 μm, and 1.8μm between *p1* and *p3*.

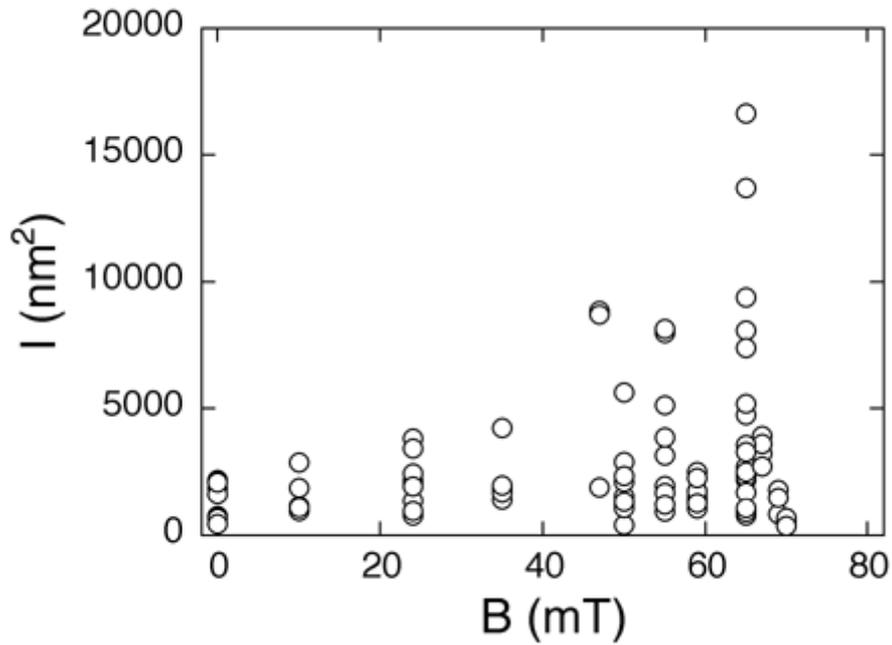

Figure 3: Intensity I ($n$m$^2$) of the cross-correlation function of the fluctuating parallel displacements of two actual reference points *p1* and *p2* (see Fig. 1) attached on a speckled actin filament at no time delay parameterized in terms of the applied magnetic flux density B ($m$T). The direction of the sliding movement for each sample was in parallel to the planar plane on which the magnetic flux was fixed, but was taken arbitrary to the linear direction of the applied magnetic flux.

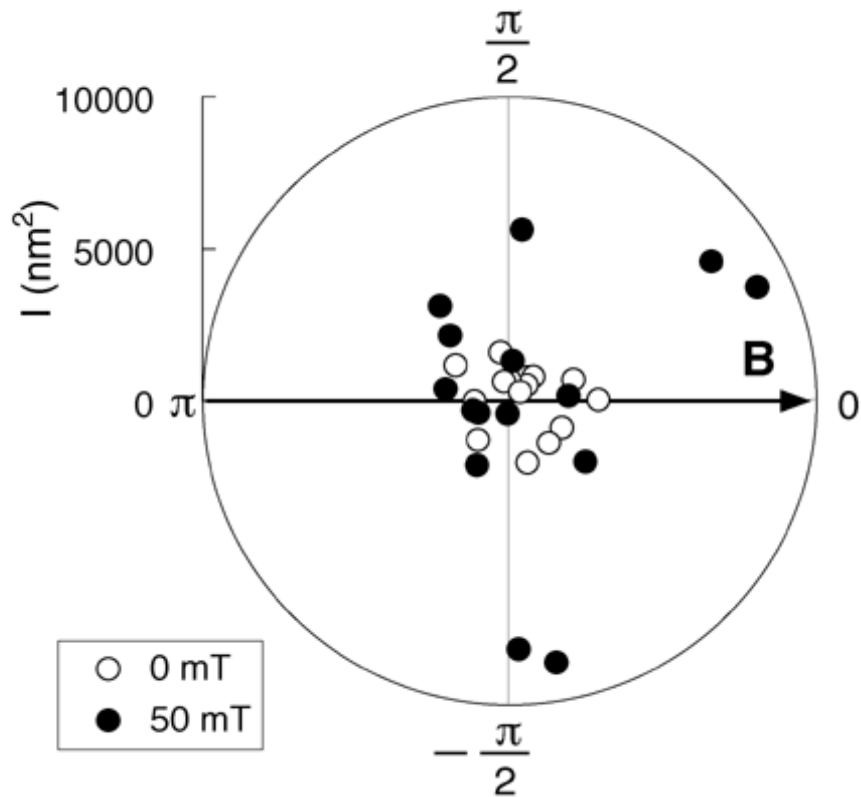

Figure 4: Intensity I ($nm^2$) of the cross-correlation function of the fluctuating parallel displacements of two actual reference points *p1* and *p2* (see Fig. 1) at no time delay parameterized in terms of the direction and the strength of the applied magnetic flux density B (*m*T). The direction of the sliding movement of the filament was measured relative to the direction towards which the magnetic flux was applied.